\documentclass[12pt]{article}
\usepackage[dvips]{color}
\usepackage{epsfig}
\usepackage{amsmath}
\usepackage{graphicx}
\def\Box{\hbox{$\rlap{$\sqcup$}\sqcap$}}
\begin{document}
\setcounter{page}{1}

\pagestyle{plain} \vspace{1cm}

\begin{center}
\Large{\bf Crossing of $\omega=-1$ with Tachyon and Non-minimal Derivative Coupling}\\
\small \vspace{1cm} {\bf A. Banijamali $^{a}$
\footnote{a.banijamali@nit.ac.ir}} and {\bf B. Fazlpour $^{b}$
\footnote{b.fazlpour@umz.ac.ir}}\\
\vspace{0.5cm}  $^{a}$ {\it Department of Basic Sciences, Babol
University of Technology, Babol, Iran\\} \vspace{0.5cm} $^{b}$ {\it
Department of Physics, Ayatollah Amoli Branch, Islamic Azad
University, P. O. Box 678, Amol, Iran\\}
\end{center}
\vspace{1.5cm}
\begin{abstract}
We construct a single scalar field model with tachyon field
non-minimally coupled to itself, its derivative and to the
curvature. We study the cosmological dynamics of the equation of
state in this setup. While it is expected that in the case of single
scalar field the crossing of the phantom divide line can not be
realized [10], we show that incorporating quantum corrections
namely, non-minimal derivative coupling of scalar field with
curvature in our model, lead to phantom divide crossing.\\


\end{abstract}
\newpage
\section{Introduction}
Recent cosmological observations have revealed that the present
state of the universe is undergoing an accelerated expansion [1-4].
This acceleration is triggered by more than 70$\,_{\circ} /^{\circ}$
of dark energy. Dark energy (DE) has been one of most active field
in modern cosmology [5]. The simplest candidate  for DE is a tiny
positive time-independent cosmological constant $\Lambda$. However,
it has two problems: 1) fine tuning or why the cosmological constant
is about 120 orders of magnitude smaller than its natural
expectation (the Planck energy density), and 2) coincidence problem
or why are we living in an epoch in which the DE density and the
dust matter energy are comparable?\\
As a possible solution to these problems, many dynamical scalar
field models of DE have been proposed. Quintessence, phantom,
k-essence and tachyon scalar fields belong to these sort of DE
models (for review see [6]).\\
In the other hand, various observational data such as SNe Ia Gold
dataset [7] confirmed that the effective equation of state (EoS)
parameter $\omega$ (the ratio of the effective pressure of the
universe to the effective energy density of it) crosses $-1$,
namely, the cosmological constant barrier, currently or in the past.
It has been shown [8-11] that with a single fluid or a single
minimally coupled scalar field it is impossible to realize EoS
crossing $-1$ and one needs to introduce extra degree of freedom to
the ordinary theories
of these kinds.\\
A number of attempts to realize the crossing of the cosmological
constant barrier are as follows: hybrid model which is composed of
two scalar fields (quintessence and phantom) [11] or three scalar
fields [12], scalar field model with non-linear kinetic terms [13]
or a non-linear higher-derivative one [10], braneworld models [14],
phantom coupled to dark matter with an appropriate coupling [15],
string inspired models [16], non-local gravity [17], modified
gravity models [18] and also non-minimally coupled scalar field
models in which scalar field couples with scalar curvature,
Gauss-Bonnet invariant or modified $f(R)$ gravity [19-21] (for a
detailed review, see [22]). Crossing of the phantom divide can also
be realized with single imperfect fluid [23] or by a constrained
single degree of freedom
dust like fluids [24].\\
Furthermore, non-minimal couplings are generated by quantum
corrections to the scalar field theory and they are essential for
the renormalizability of the scalar field theory in curved space
(see [25] and references therein). One can extend the non-minimally
coupled scalar tensor theories, allowing for non-minimal coupling
between the derivatives of the scalar fields and the curvature [26].
A model with non-minimal derivative coupling was proposed in [26-28]
and interesting cosmological behaviors of such a model in
inflationary cosmology [29], quintessence and phantom cosmology [30,
31], asymptotic solutions and restrictions on the coupling parameter
[32] have been widely studied in the literature. General non-minimal
coupling of a scalar field and kinetic term to the curvature as a
source of DE has been analyzed in [33]. Also, non-minimal coupling
of modified gravity and kinetic part of Lagrangian of a massless
scalar field has been investigated in [34]. It has been shown that
inflation and late-time cosmic acceleration of the universe can be
realized in such a model. \\
In this paper we consider an explicit coupling between the scalar
field, the derivative of the scalar field and the curvature and
study crossing of the $\omega=-1$ in such a model. We are interested
in our analysis to the case of tachyon scalar field. The tachyon
field in the world volume theory of the open string stretched
between a D-brane and an anti-D-brane or a non-BPS D-brane plays the
role of scalar field in the context of string theory [35]. What
distinguishes the tachyon Lagrangian from the standard Klein-Gordan
form for scalar field is that the tachyon action has a non-standard
type namely, Dirac-Born-Infeld form [36]. Moreover, the tachyon
potential is derived from string theory and should be satisfy some
definite properties to describe tachyon condensation and other
requirements in string theory. In summary, our motivation for
investigating a model with non-minimal derivative coupling and
tachyon scalar field is coming from a fundamental theory such as
string/superstring theory and it may provide a possible approach to
quantum gravity from a perturbative point of view [37-39].\\
An outline of the present work is as follows: In section 2 we
introduce a model of DE in which the tachyon field plays the role of
scalar field and the non-minimal coupling between scalar field, the
time derivative of scalar field and Einstein tensor is also present
in the action. Then we derive field equations as well as energy
density and pressure of the model in order to study the EoS
parameter behavior in section 3. We obtain the conditions required
for $\omega$ crossing $-1$ and using numerical method, we will show
that the model can realize the $\omega=-1$ crossing. Section 4
is devoted to our conclusions.\\

\section{Field Equations}
We consider the following Born-Infeld type action for tachyon field
with non-minimal derivative coupling and also with itself,
\begin{equation}
S=\int d^{4}x
\sqrt{-g}\Big[\frac{1}{2k^{2}}R-V(\phi)\sqrt{1+g^{\mu\nu}\partial_{\mu}\phi\partial^{\nu}\phi}+\xi
f(\phi)G_{\mu\nu}\partial^{\mu}\phi\partial^{\nu}\phi\Big],
\end{equation}
where $\kappa^{2} = 8\pi G = \frac{1}{M_{Pl}^{2}}$ while $G$ is a
bare gravitational constant and $M_{Pl}$ is a reduced Planck mass,
$V(\phi)$ is the tachyon potential which is bounded and reaching its
minimum asymptotically. $f(\phi)$ is a general function of the
tachyon field $\phi$ and $\xi$ is coupling constant. The models of
kind (1) with non-minimal coupling between derivatives of a scalar
field and curvature are the extension of scalar-tensor theories.
Such a non-minimal coupling may appear in some Kaluza-Klein theories
[40, 41]. In Ref. [26], Amendola has considered a model with
non-minimal coupling between derivative of scalar field and the
Ricci scalar, $\xi R\partial_{\mu}\phi\partial^{\mu}\phi$, and by
using generalized slow-roll approximations, he has obtained some
inflationary solutions of this model.\\
A general model containing two derivative coupling terms $\xi_{1} R
\partial_{\mu}\phi \partial^{\mu}\phi$ and $\xi_{2} R_{\mu\nu} \partial^{\mu}\phi
\partial^{\nu}\phi$, has been studied in [27, 28]. It was shown in [30]
that field equations of this theory are of third order in
$g_{\mu\nu}$ and $\phi$, but in the special case where
$-2\xi_{1}=\xi_{2}=\xi$ the order of equations are reduced to the
second order. This particular choice of $\xi_{1}$ and $\xi_{2}$
leads to the non-minimal coupling between derivative of scalar field
and the Einstein tensor, $\xi G_{\mu\nu}
\partial^{\mu}\phi\partial^{\nu}\phi$. Sushkov in [30] has obtained
the exact cosmological solutions of this theory and he has concluded
that such a model is able to explain a quasi-de sitter phase as
well as an exit from it without any fine-tuned potential.\\
Varying the action (1) with respect to metric tensor $g_{\mu\nu}$,
leads to
\begin{equation}
G_{\mu\nu}=R_{\mu\nu}-\frac{1}{2}g_{\mu\nu}R=k^{2}\big(T_{\mu\nu}+T'_{\mu\nu}\big),
\end{equation}
where
\begin{equation}
T_{\mu\nu}=\nabla_{\mu}\phi\nabla_{\nu}\phi-\frac{1}{2}g_{\mu\nu}\big(\nabla\phi\big)^{2}-g_{\mu\nu}V(\phi),
\end{equation}
and
$$T'_{\mu\nu}=R\big(\nabla_{\mu}\phi\nabla_{\nu}\phi\big)-4\nabla_{\gamma}\phi\nabla_{(\mu}\phi
R^{\gamma}_{\nu)}+G_{\mu\nu}\big(\nabla\phi\big)^{2}-2R_{\mu\nu\gamma\lambda}\nabla^{\gamma}
\phi\nabla^{\lambda}\phi-2\nabla_{\mu}\nabla^{\gamma}\phi\nabla_{\nu}\nabla_{\gamma}\phi$$
\begin{equation}
+2\nabla_{\mu}
\nabla_{\nu}\phi\Box\phi+g_{\mu\nu}\Big(\nabla^{\gamma}\nabla^{\lambda}\phi\nabla_{\gamma}\nabla_{\lambda}
\phi-\big(\Box\phi\big)^{2}+2R^{\gamma\lambda}\nabla_{\gamma}\phi\nabla_{\lambda}\phi\Big).
\end{equation}
Scalar field equation of motion can be obtain by varying (1) with
respect to $\phi$,
\begin{equation}
\nabla_{\mu}\Big(\frac{V(\phi)\nabla^{\mu}\phi}{u}\Big)-\frac{dV(\phi)}{d\phi}u-\xi
f(\phi) G^{\mu\nu}\nabla_{\mu}\nabla_{\nu}\phi+\xi
\frac{df(\phi)}{d\phi}G_{\mu\nu}\partial^{\mu}\phi\partial^{\nu}\phi=0,
\end{equation}
where $u=\sqrt{1+\nabla_{\mu}\phi\nabla^{\mu}\phi}$.\\
For a spatially-flat Friedmann-Robertson-Walker (FRW) metric,
\begin{eqnarray}
ds^{2}=-dt^{2}+a^{2}(t)(dr^{2}+r^{2}d\Omega^{2}),
\end{eqnarray}
the  components of the Ricci tensor $R_{\mu\nu}$ and the Ricci
scalar $R$ are given by
\begin{equation}
R_{00}=-3\big(\dot{H}+H^{2}\big),\,\,R_{ij}=a^{2}(t)\big(\dot{H}+3H^{2}\big)\delta_{ij},\,\,
R=6\big(\dot{H}+2H^{2}\big),
\end{equation}
where $H=\frac{\dot{a}(t)}{a(t)}$ is the Hubble parameter and $a(t)$
is the scale factor. The equation of motion of the scalar field for
a homogeneous $\phi$ in FRW background (6) takes the following form
$$\frac{\ddot{\phi}}{1-\dot{\phi}^{2}}+3H\dot{\phi}+\frac{1}{V(\phi)}\frac{dV}{d\phi}$$
\begin{equation}
+\frac{\sqrt{1-\dot{\phi}^{2}}}{V(\phi)} \Bigg(3\xi
H^{2}\Big(2f(\phi)\ddot{\phi}+\frac{df}{d\phi}\dot{\phi}^{2}\Big)+18\xi
H^{3}f(\phi)\dot{\phi}+12\xi H \dot{H} f(\phi)\dot{\phi}\Bigg)=0.
\end{equation}
The $(0,0)$ component and $(i,i)$ components of equation (2)
correspond to energy density and pressure respectively,
\begin{equation}
\rho=\frac{V(\phi)}{\sqrt{1-\dot{\phi}^{2}}}+9\xi
H^{2}f(\phi)\dot{\phi}^{2},
\end{equation}
and
\begin{equation}
P=-V(\phi)\sqrt{1-\dot{\phi}^{2}}-\xi\big(3H^{2}+2\dot{H}\big)f(\phi)\dot{\phi}^{2}-2\xi
H\Big(2f(\phi)\dot{\phi}\ddot{\phi}+\frac{df}{d\phi}\dot{\phi}^{3}\Big).
\end{equation}
Friedmann equation is also as follows,
\begin{equation}
H^{2}=\frac{\kappa^{2}}{3}\Big(\frac{V(\phi)}{\sqrt{1-\dot{\phi}^{2}}}+9\xi
H^{2}f(\phi)\dot{\phi}^{2}\Big).
\end{equation}
Next, we want to investigate the effects of non-minimal derivative
coupling on the cosmological evolution of EoS and see how the
present model can be used to realize a crossing of phantom divide
$\omega=-1$.

\section{The $\omega=-1$ Crossing with Tachyon Field}
To study the cosmological consequence of the present model we start
with $\omega=\frac{P}{\rho}$. From the definition of EoS one can
write $P+\rho=(1+\omega)\rho$. Using equations (9) and (10) we have
the following expression,
\begin{equation}
\rho+P=\frac{V(\phi) \dot{\phi}^{2}}{\sqrt{1-\dot{\phi}^{2}}}+6\xi
H^{2}f(\phi)\dot{\phi}^{2}-2\xi\dot{H}f(\phi)\dot{\phi}^{2}-2\xi
H\Big(2f(\phi)\dot{\phi}\ddot{\phi}+\frac{df}{d\phi}\dot{\phi}^{3}\Big).
\end{equation}
The above equation must be zero when $\omega\rightarrow -1$. In
order to achieve this requirement, we obtain two following
possibilities,\\
\begin{equation}
\dot{\phi}=0,
\end{equation}
or
\begin{equation}
\dot{\phi}\Big(\frac{V(\phi)}{\sqrt{1-\dot{\phi}^{2}}}+6\xi
H^{2}f(\phi)-2\xi\dot{H}f(\phi)\Big)=2\xi
H\Big(2f(\phi)\ddot{\phi}+\frac{df}{d\phi}\dot{\phi}^{2}\Big).
\end{equation}
Also, we have to check $\frac{d}{dt}(\rho+P)\neq 0$, when $\omega$
crosses over $-1$,
$$\frac{d}{dt}(\rho+P)=\frac{V(\phi)
\dot{\phi}^{3}}{\sqrt{1-\dot{\phi}^{2}}}+\frac{2V(\phi)
\dot{\phi}\ddot{\phi}}{\sqrt{1-\dot{\phi}^{2}}}+\frac{V(\phi)
\dot{\phi}^{3}\ddot{\phi}}{\big(1-\dot{\phi}^{2}\big)^{\frac{3}{2}}}+
2\xi\big(3
H^{2}-2\dot{H}\big)\Big(2f(\phi)\dot{\phi}\ddot{\phi}+\frac{df}{d\phi}\dot{\phi}^{3}\Big)$$
\begin{equation}
+2\xi\big(6H \dot{H}-\ddot{H}\big)f(\phi)\dot{\phi}^{2}-2\xi
H\Big(2f(\phi)\big(\dot{\phi}\dddot{\phi}+\ddot{\phi}^{2}\big)+5\frac{d
f}{d\phi}\dot{\phi}^{2}\ddot{\phi}+\frac{d^{2}f}{d\phi^{2}}\dot{\phi}^{4}\Big).
\end{equation}
If we assume the first case, namely $\dot{\phi}=0$, when $\omega$
crosses the phantom divide line, then equation (15) can be rewritten
as the following form
\begin{equation}
\frac{d}{dt}(\rho+P)=-4\xi H f(\phi)\ddot{\phi}^{2}.
\end{equation}
It is clear from (16) that, the additional condition for having
crossing of the phantom divide is $\Rightarrow\ddot{\phi}\neq0$. One
concludes from the above discussion that crossing of the phantom
divide line in our model must be happen before reaching potential to
its minimum, because at the minimum of the tachyon potential, we
have $\ddot{\phi}=0$ and $\dot{\phi}\neq 0$ and this is a well known
property of $V(\phi)$. Therefore when $\omega$ crosses $-1$ the
tachyon field should continue to run away
since $\ddot{\phi}\neq0$.\\
The Authors of Ref. [42] have obtained the same result as ours. But
in their proposal, they have inserted an extra term, $\phi\Box\phi$,
in square root part of tachyon Lagrangian by hand. Note that in our
model there is no extra term but we have included non-minimal
coupling of tachyon field with its derivative and curvature due to
quantum corrections.\\
In the next step we consider the second possibility in equation
(14). Then equation (15) takes the following form
$$\frac{d}{dt}(\rho+P)=\frac{V(\phi)
\dot{\phi}^{3}}{\sqrt{1-\dot{\phi}^{2}}}\Big(1+\frac{\ddot{\phi}}{1-\dot{\phi}^{2}}\Big)
-2\xi\ddot{H}f(\phi)\dot{\phi}^{2}+6\xi
H^{2}\frac{df}{d\phi}\dot{\phi}^{3}-2\xi
H\Big(2f(\phi)\dot{\phi}\dddot{\phi}-2f(\phi)\ddot{\phi}^{2}$$
\begin{equation}
+3\frac{d f}{d\phi}\dot{\phi}^{2}\ddot{\phi}+
\frac{d^{2}f}{d\phi^{2}}\dot{\phi}^{4}\Big).
\end{equation}
We can see that even if $\ddot{\phi}=0$ and $\dddot{\phi}=0$,
crossing $-1$ can be happen. In this case our results are the same
as those obtained in Ref. [21] where tachyon field non-minimally
coupled to Gauss-Bonnet invariant. So, it seems that in studying
phantom divide crossing cosmology the non-minimal coupling of
tachyon field with this derivative and Einstein tensor has the same
effects as coupling of tachyon to Gauss-Bonnet invariant where
crossing over $-1$ can be
happen when tachyon potential reaches its minimum asymptotically.\\
In order to show that our model can realize crossing of $\omega=-1$
more clearly, we choose two specific tachyon potential and study
evolution of EoS numerically. Figure 1 shows such a numerical
calculations for $V(\phi)=V_{0}e^{-\alpha \phi^{2}}$ with constant
$\alpha$. One can see that the model predicts crossing of $-1$ at
redshift $z=1.65$. In figure 2 we have taken another tachyon
potential $V(\phi)=\frac{V_{0}}{\phi^{2}}$. It has been shown that
crossing of $\omega=-1$ can be realized in our model. In this case
crossing of phantom divide takes place at $z=1.32$. Also we have
used the function $f(\phi)=b\phi^{n}$ with constants $b$ and $n$.\\
Finally we note the following points: in our numerical calculations
if we do not consider the non-minimal coupling of scalar field with
itself, namely $f(\phi)=1$ [30, 31], then for
$V(\phi)=V_{0}e^{-\alpha \phi^{2}}$ crossing of $\omega=-1$ can not
be realized and for $V(\phi)=\frac{V_{0}}{\phi^{2}}$ the EoS will be
a constant larger
than $-1$ hence it doesn't cross the phantom divide.\\
\begin{figure}[htp]
\begin{center}
\includegraphics{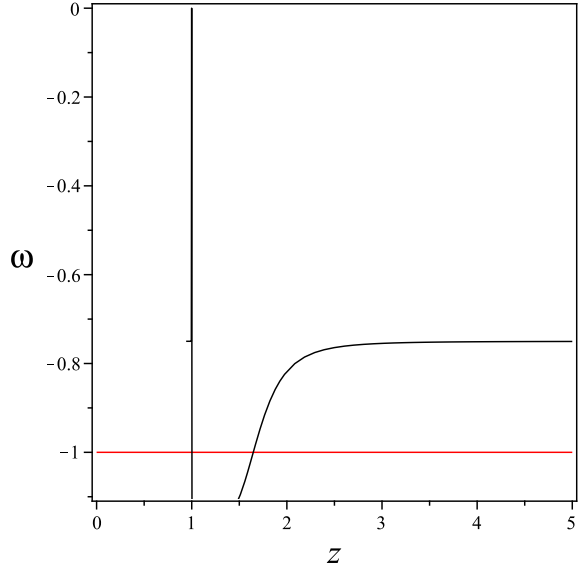}
\end{center}
\vspace{6 cm}
 \caption{\small { The plot
of EoS parameter versus redshift $z$ for the potential
 $V(\phi)=V_{0}e^{-\alpha \phi^{2}}$, $\phi=\phi_{0}t$, $f(\phi)=b\phi^{n}$ and $H=\frac{h_{0}}{t}$, (with $\xi=10$, $b=1$,
 $n=8$, $V_{0}=4$, $h_{0}=100$, $\phi_{0}=0.5$ and $\alpha=5$). Crossing takes place at $z=1.65$.}}
\end{figure}
\begin{figure}[htp]
\begin{center}
\includegraphics{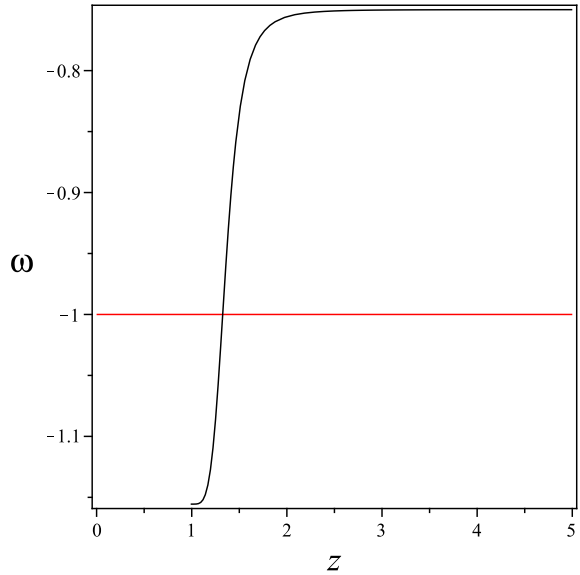}
\end{center}
\vspace{5 cm}
 \caption{\small {The plot
of EoS parameter versus redshift $z$ for the potential
 $V(\phi)=\frac{V_{0}}{\phi^{2}}$, $\phi=\phi_{0}t$, $f(\phi)=b\phi^{n}$ and $H=\frac{h_{0}}{t}$, (with $\xi=10$, $b=1$,
 $n=8$, $V_{0}=4$, $h_{0}=100$ and $\phi_{0}=0.5$). Crossing takes place at $z=1.32$.}}
\end{figure}\\

\newpage
\section{Conclusion}
We have considered the gravitational theory of a scalar field with
non-minimal derivative coupling to curvature and itself. We have
studied cosmological evolution of EoS in this setup where tachyon
field played the role of scalar field. We have shown that there are
two possibilities to have phantom divide crossing to such a model.
These possibilities are given by equations (13) and (14). By
choosing the condition (13), we have concluded that the crossing
over $-1$ must be happen before reaching tachyon potential to its
minimum and this is the same result as that in [42]. In the other
side if we consider the second possibility namely, the condition
(14), it has been shown that the $\omega=-1$ crossing can be
realized even if the potential goes to its minimum asymptotically.
Our result in this case is the same as [21]. We have also
investigated our model numerically and showed that the crossing of
phantom divide occur for special potentials and coupling function.
It may be interesting to consider different potentials and coupling
functions in this setup.\\

\end{document}